\documentclass[12pt]{article}
\usepackage{graphicx}
\RequirePackage{xspace}
\usepackage{relsize}

\def\pbnr{}
\def\speaker{Gianluca Inguglia}
\def\onbehalfof{}
\def\title{Escaping from controversies in $CP$ violation measurements in charm decays}
\def\affiliation{School of Physics and Astronomy\\
Queen Mary, University of London, London, UK}
\def\support{The workshop was supported by the University of Manchester, IPPP, STFC, and IOP}

\newcommand\pubnumber{\pbnr}
\newcommand\pubdate{\today}
\def\Title#1{\begin{center} {\Large #1 } \end{center}}
\def\Author#1{\begin{center}{ \sc #1} \end{center}}

\newcommand{\OnBehalf}[1]{\sbox0{#1}\ifdim\wd0=0pt
        {}
	\else
	{\\on behalf of #1}
	\fi}
\newcommand{\SupportedBy}[1]{\sbox0{#1}\ifdim\wd0=0pt
        {}
	\else
	{\footnote{#1}}
	\fi}
\def\Address#1{\begin{center}{ \it #1} \end{center}}

\newcommand\pubblock{\includegraphics[width=5cm]{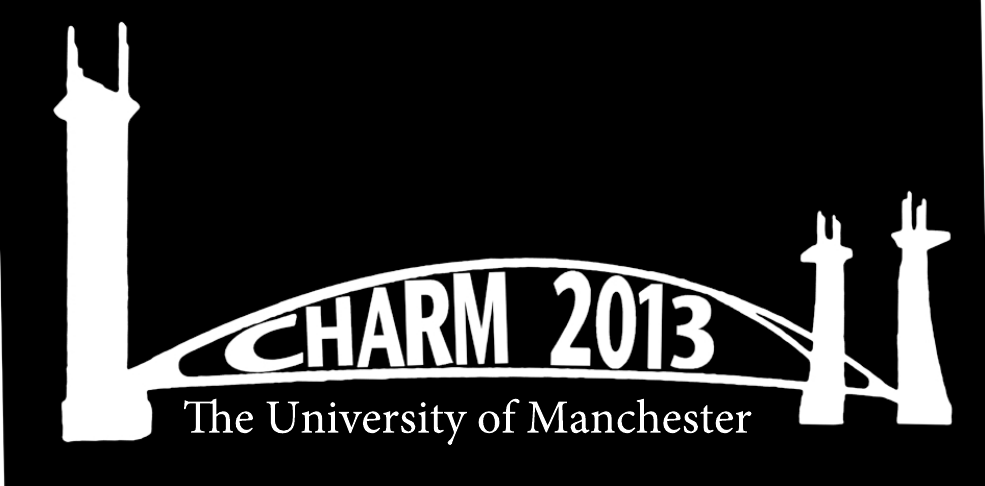}\hfill{\begin{tabular}{l} \pubnumber\\
         \pubdate  \end{tabular}}}
\newenvironment{Abstract}{\begin{quotation}  }{\end{quotation}}
\newenvironment{Presented}{\begin{quotation} \begin{center} 
             PRESENTED AT\end{center}\bigskip 
      \begin{center}\begin{large}}{\end{large}\end{center} \end{quotation}}
\def\Acknowledgements{\bigskip  \bigskip \begin{center} \begin{large}
             \bf ACKNOWLEDGEMENTS \end{large}\end{center}}
\def\venue{The 6$^{th}$ International Workshop on Charm Physics\\
(CHARM 2013)\\
Manchester, UK,  31 August -- 4 September, 2013}

\def\beq{\begin{equation}}
\def\eeq#1{\label{#1}\end{equation}}
\def\eeqn{\end{equation}}

\def\beqa{\begin{eqnarray}}
\def\eeqa#1{\label{#1}\end{eqnarray}}
\def\eeqan{\end{eqnarray}}

\let\bar=\overbar

\def\Dslash{\not{\hbox{\kern-4pt $D$}}}
\def\dslash{\not{\hbox{\kern-2pt $\del$}}}

\def\msb{{\bar{\ssstyle M \kern -1pt S}}}

\begin{document}
\begin{titlepage}
\pubblock

\vfill
\Title{\title}
\vfill
\Author{\speaker\SupportedBy{\support}\OnBehalf{\onbehalfof}}
\Address{\affiliation}
\vfill
\begin{Abstract}
The breaking of the $CP$ symmetry in $D^0$ meson decays has been awaited for a long time. After a set of measurements provided by the LHCb, CDF, and Belle Collaborations leading in march 2012 to combined results that were consistent with no $CP$ violation at a CL of $0.006\%$ suggesting $CP$ violation at $\sim 1\%$ level. Such a potentially large value of $CP$ violation in charm decays has triggered widespread interest from the whole particle physics community to evaluate the implications of such an interesting unexpected results. However, a more recent combination of more up-to-date results in March 2013, has slightly changed the situation, showing that data are consistent with the $CP$ conserving hypothesis at $2.1\%$ CL. I briefly review the method used by the various Collaborations when extracting the quantity $\Delta A_{CP}$ and the relative results. Finally I discuss the need for additional measurements, and present the potential of a time-dependent analysis when looking for $CP$ violation in $D^0$ decays and how this can be used to largely improve the current sensitivity on the mixing phase $\phi_{MIX}$.

\end{Abstract}
\vfill
\begin{Presented}
\venue
\end{Presented}
\vfill
\end{titlepage}
\def\thefootnote{\fnsymbol{footnote}}
\setcounter{footnote}{0}
%
\def\hbabar{\mbox{{\huge\bf\sl B}\hspace{-0.1em}{\LARGE\bf\sl A}\hspace{-0.03em}{\huge\bf\sl B}\hspace{-0.1em}{\LARGE\bf\sl A\hspace{-0.03em}R}}\xspace}
\def\Lbabar{\mbox{{\LARGE\sl B}\hspace{-0.15em}{\Large\sl A}\hspace{-0.07em}{\LARGE\sl B}\hspace{-0.15em}{\Large\sl A\hspace{-0.02em}R}}\xspace}
\def\lbabar{\mbox{{\large\sl B}\hspace{-0.4em} {\normalsize\sl A}\hspace{-0.03em}{\large\sl B}\hspace{-0.4em} {\normalsize\sl A\hspace{-0.02em}R}}\xspace}
\def\babar{\mbox{\slshape B\kern-0.1em{\smaller A}\kern-0.1em
    B\kern-0.1em{\smaller A\kern-0.2em R}}\xspace}
\section{Time integrated \boldmath{$CP$} test in \boldmath{$D^0 \to h^+h^- ~(h=K,\pi)$}: \boldmath{$\Delta A_{CP}$}}

\begin{figure}[!ht]
\begin{center}
\resizebox{10.cm}{!}{
\includegraphics{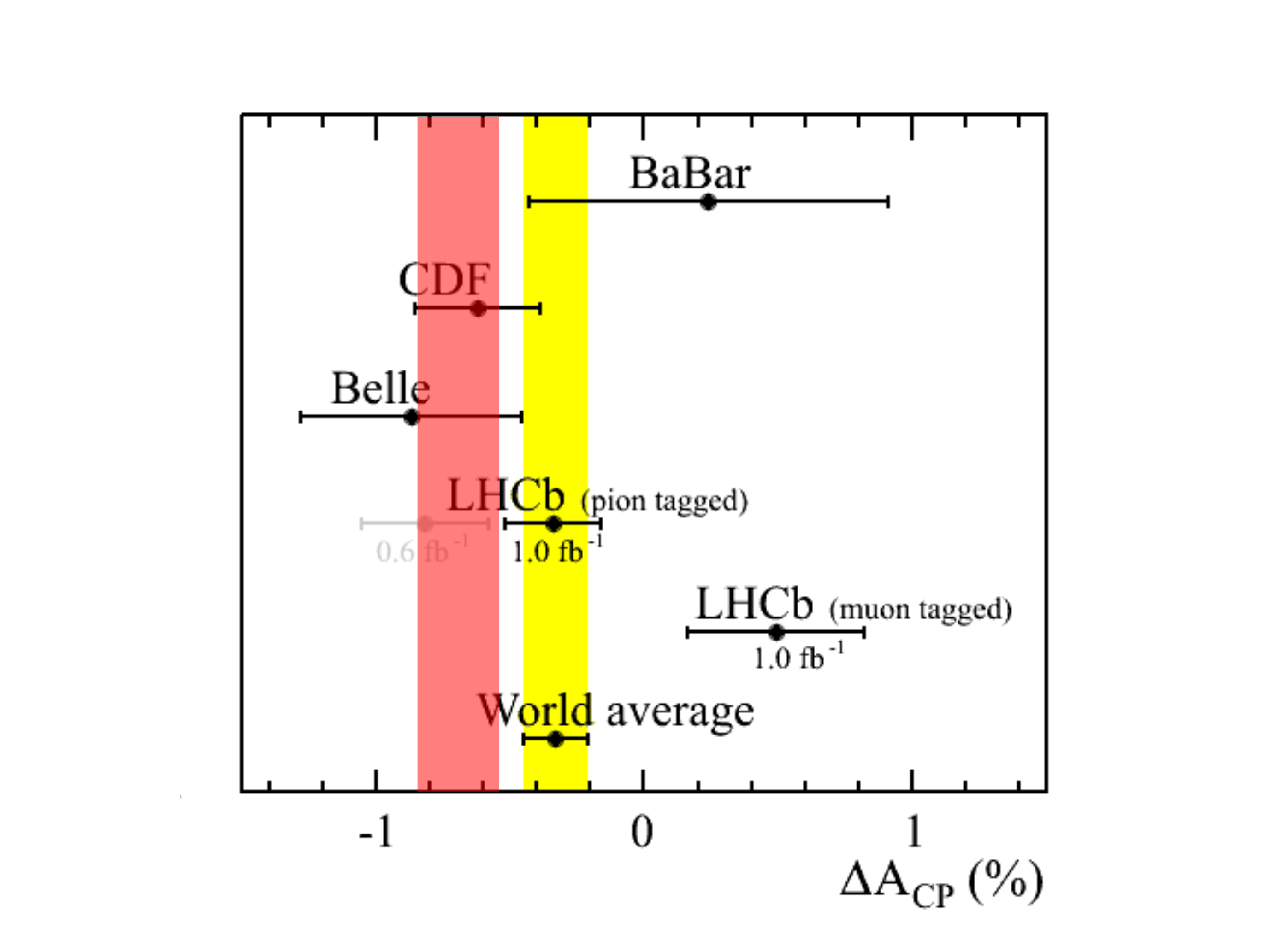}
}
\caption{Comparison of different measurements of $\Delta A_{CP}$, the yellow band indicates a naive average of the various results, the red band has been added to the original picture to show a naive average of results prior to the 2013 update.}\label{fig:combined}
\end{center}
\end{figure}

In the past few years lot of work has been done by various Collaborations to perform tests of $CP$ violation in the charm system. Different tests of $CP$ violation can be done and these are \textit{direct} (time integrated), \textit{indirect}, and in the \textit{interference between mixing and decay} (time-dependent). Most of the efforts up to now have been focused on potential measurements of time integrated $CP$ violation, for which the asymmetry for $D^0$ meson is simply given by
\begin{eqnarray}
A_{CP}(f_{CP})= \frac{|\overline{A}|^2- |A|^2}{|\overline{A}|^2 + |A|^2}\label{eq:asymmetry}
\end{eqnarray}
where $A$ and $\overline{A}$ are the amplitudes of the the decays of $D^0$ and $\overline{D}^0$ mesons to a particular $CP$ eigenstate $f_{CP}$. Experimentally one would measure the following:
\begin{eqnarray}
A_{CP}(f_{CP})= \frac{N(D^0 \to f_{CP})- N(\overline{D}^0 \to f_{CP})}{N(D^0 \to f_{CP})+ N(\overline{D}^0 \to f_{CP})}\label{eq:asymmetry1}
\end{eqnarray}
where $N$ indicates the number of observed decays. The situation unfortunately is complicated by the fact that systematic effects may play a very important role here (detector asymmetries, mistag rate, etc.).
Generally, in A hadron collider one would expect that
\begin{eqnarray}
A_{CP}(f_{CP})= A_{raw}(f_{CP})-A_D(f_{CP})-A_D(\pi_s^{\pm})-A_P(D^{*\pm}) \label{eq:lhcbraw}
\end{eqnarray}
where $A_{raw}(f_{CP})$ is the observed asymmetry, $A_D(f_{CP})$ is the asymmetry coming from the selection of $D$ decaying to $f_{CP}$, $A_D(\pi_S^{\pm})$ is the asymmetry coming from the selection of $\pi_s^{\pm}$ and $A_P(D^{*\pm})$ is the production asymmetry for $D^{*\pm}$.
And similarly for an $e^+e^-$ collider running at the centre-of-mass energy equivalent to that of the $\Upsilon(4S)$ for example, one would expect:
\begin{eqnarray}
A_{CP}(f_{CP})= A_{raw}(f_{CP})-A_{FB}-A_\epsilon(f_{CP}) \label{eq:belleraw}
\end{eqnarray} 
where $A_{FB}$ is the \textit{forward-backward} asymmetry for $e^+e^-\to c \overline{c}$ processes and $A_\epsilon(f_{CP})$ is the particle detection asymmetry which depends on the final state and can be calculated by using large samples of $CP$ conserving decays~\cite{belle}.
The various systematic effects make the measurement rather difficult, nonetheless the CDF, Belle, and \babar Collaborations have performed the analysis for $D^0 \to h^+h^-$ $(h=K,\pi)$~\cite{cdf}~\cite{belle1}~\cite{babar}, finding 
\begin{eqnarray}
CDF:~~
A_{CP}(D^0 \to \pi^+\pi^-)&=& +0.31 \pm 0.22 (\%)\nonumber \\ 
A_{CP}(D^0 \to K^+ K^-)&=& -0.32 \pm 0.21 (\%)\label{eq:CDF}
\end{eqnarray} 
where the errors include both statistical and systematic effects, showing an agreement with the $CP$ conserving hypothesis at $1.5 \sigma$, and 
\begin{eqnarray}
Belle: ~~
A_{CP}(D^0 \to \pi^+\pi^-)&=& (+0.55 \pm 0.36_{stat} \pm 0.09_{sys}) (\%) \nonumber \\ 
A_{CP}(D^0 \to K^+ K^-)&=& (-0.32 \pm 0.21_{stat}  \pm 0.09_{sys}) (\%)\label{eq:belle}
\end{eqnarray} 
\begin{eqnarray}
\babar: ~~
A_{CP}(D^0 \to \pi^+\pi^-)&=& (-0.24 \pm 0.52_{stat} \pm 0.22_{sys}) (\%) \nonumber \\ 
A_{CP}(D^0 \to K^+ K^-)&=&~~  (0.00 \pm 0.34_{stat}  \pm 0.13_{sys})(\%).\label{eq:belle}
\end{eqnarray} 
An additional possibility to measure $CP$ violation in the two discussed decays consists in taking their combination. Due to the particular kind of systematic effects affecting the two single measurements one can take the difference, resulting in
\begin{eqnarray}
\Delta A_{raw} = A_{CP}(K^+K^-)+A_D(K^+K^-)+A_D(\pi_s^{\pm})+A_P(D^{*\pm}) &-& \nonumber  \\ A_{CP}(\pi^+\pi^-)+A_D(\pi^+\pi^-)+A_D(\pi_s^{\pm})+A_P(D^{*\pm}) &=& \nonumber \\ A_{CP}(K^+K^-)-A_{CP}(\pi^+\pi^-)+\Delta A_{\pi^+\pi^-}^{K^+K^-} \label{eq:raw1}
\end{eqnarray}
where $\Delta A_{\pi^+\pi^-}^{K^+K^-}=A_D(K^+K^-)-A_D(\pi^+\pi^-)$. However, as a result of the symmetry inherent in the decay of a $D^0$ meson into two spin zero $CP$ conjugate particles, there is no detection asymmetry, i.e. $A_D(K^+K^-)=A_D(\pi^+\pi^-)=0$~\cite{lhcbold}. Hence
\begin{eqnarray}
\Delta A_{raw}=A_{CP}(K^+K^-)-A_{CP}(\pi^+\pi^-)=\Delta A_{CP}\label{eq:rawcp}
\end{eqnarray}
showing that the extracted value of $\Delta A_{raw}$ is as a measurement of $CP$ violation. The Belle~\cite{belle1}, CDF~\cite{cdf}, and LHCb~\cite{lhcbold}~\cite{lhcbmuon}~\cite{lhcbnew} Collaborations have performed searches of $CP$ violation in charm decays by combining the $D^0 \to K^+K^-$ and $D^0 \to \pi^+\pi^-$ channels, finding:
\begin{eqnarray}
\mathrm{Belle} (2012): \Delta A_{CP} &=& -0.87\pm 0.41_{stat} \pm 0.06_{sys} (\%)\\
\mathrm{CDF}   (2012): \Delta A_{CP} &=& -0.62\pm 0.21_{stat} \pm 0.10_{sys} (\%)\\ 
\mathrm{LHCb}  (0.6 fb^{-1}, prompt~D^*)(2012):\Delta A_{CP} &=& -0.82\pm 0.21_{stat} \pm 0.11_{sys} (\%)\\
\mathrm{LHCb}  (1.0 fb^{-1}, prompt~D^*)(2013):\Delta A_{CP} &=& -0.34\pm 0.15_{stat} \pm 0.10_{sys} (\%)\\
\mathrm{LHCb}  (1.0 fb^{-1}, semilep.~b)(2013):\Delta A_{CP} &=& +0.49\pm 0.30_{stat} \pm 0.14_{sys} (\%)\label{eq:CDF}
\end{eqnarray} 
where one can easily see that the results from Belle, CDF, and LHCb up to 2012 were in good agreement with a large value of $\Delta A_{CP}$ ($\Delta A_{CP}^{dir}=-0.678 \pm 0.147(\%)$), while 2013 measurements push the value of $\Delta A_{CP}$ toward zero ($\Delta A_{CP}^{dir}=-0.329 \pm 0.121(\%)$)~\cite{hfag}, and in particular the measurement performed using $D^0$ mesons coming from inclusive semileptonic $b$-hadron decays results in a surprisingly positive value of $\Delta A_{CP}$.\footnote{The result is surprising because $\Delta A_{CP}$ should not depend on the tagging mode used to identify the flavour of the $D^0/\overline{D}^0$ mesons, in fact $CP$ is violated in the decays of a given meson and not in the production. It would be interesting to see results using higher statistics from the LHC as well as from control sample cross checks such as $D^0 \to K\pi$ and charge conjugated events coming from semileptonic decays of a $B$ meson.} The situation is depicted in Fig.~(\ref{fig:combined})~\cite{lhcbpic}, where additional measurements have been taken into account to produce a 
naive average of $\Delta A_{CP}$ (yellow band), and where a red band has been added showing a naive average of 2012 results. It has to be mentioned that these measurements are not a complete test of $CP$ violation in the charm sector, but these are rather a first step towards the determination of $CP$ violating parameters. In fact when writing Eq.~(\ref{eq:asymmetry}) the decay amplitudes have been considered, but an explicit expression for these has been omitted. For simplicity one may assume that there are two interfering amplitudes with different strong and weak phases that contribute to the decays considered (ie. $D^0 \to h^+h^-,h=K,\pi$) so that 
\begin{eqnarray}
A=A_N e^{i(\phi_N+\delta_N)}+A_T e^{i(\phi_T+\delta_T)}\label{eq:amplitudephase}
\end{eqnarray}
where a \textit{new} amplitude $A_N$ (here the distinction between a standard model penguin amplitude or a new physics contribution is considered arbitrary, and e are only concerned with the next to leading contribution) with strong and weak phases respectively $\delta_N$ and $\phi_N$ interfere with the leading \textit{tree} amplitude $A_T$ with strong and weak phases respectively $\delta_T$ and $\phi_T$. The weak and strong phase differences are $\Delta\delta$ and $\Delta\phi$, and Eq.~(\ref{eq:asymmetry}) can be written as
\begin{eqnarray}
A_{CP} = \frac{ 2 A_N A_T \sin{\Delta\phi} \sin{\Delta\delta}}{|A_N|^2+|A_T|^2+2A_NA_T\cos{\Delta\phi}\cos{\Delta\delta}}.\label{eq:asymmetry2}
\end{eqnarray} 
It can be shown~\cite{bim1} that with current measurements it is not possible to constrain $\Delta\delta$ and $\Delta\phi$, making it extremely difficult to interpret any non-zero $CP$ asymmetry (as $\Delta\phi$ is hard to calculate reliably and is a priori unknown). A possible step forward for these analysis is to consider the single decay channels separately with larger data samples, and perform independent tests of $CP$ violation using alternative methods.  
\section{Time-dependent \boldmath{$CP$} asymmetries in charm decays: potential implications}
Time-dependent measurements of $CP$ asymmetries in $D^0$ decays represent an additional tool not only to test $CP$ violation and when looking for new physics, but such measurements may provide a determination of $\phi_{MIX}$ with a higher precision than currently achieved~\cite{bim}~\cite{inguglia}. It has been shown that for un-correlated production of $D^0$ mesons the asymmetry can be written as~\cite{bim}

\begin{eqnarray}
{\cal A}^{Phys}(t) =  \Delta \omega + \frac{ (D - \Delta\omega)e^{\Delta \Gamma t/2}[ (|\lambda_{f}|^2 - 1)\cos\Delta M t + 2 Im\lambda_{f} \sin\Delta M t ]}{h_+ (1+|\lambda_{f}|^2)/2 + Re(\lambda_{f}) h_-} \label{eq:asymtagging}
\end{eqnarray}
or, in terms of $x$ and $y$ as~\cite{inguglia}

\begin{eqnarray}
\tiny
{\cal A}^{Phys}_{x,y}( t) = \Delta \omega + \frac{ (D - \Delta\omega)e^{y\Gamma t}[ (|\lambda_{f}|^2 - 1)\cos x\Gamma t + 2 Im\lambda_{f} \sin x\Gamma t ]}{h_+ (1+|\lambda_{f}|^2)/2 + Re(\lambda_{f}) h_-}
  \label{eq:asymtaggingxy}
\end{eqnarray}
where $\omega (\overline{\omega})$ represents the mistag probability for $D^0$ ($\overline{D}^0$) mesons, $\Delta\omega =\omega - \overline\omega$ and $D$ represents the dilution.
As discussed in~\cite{bim} the difference in the observed asymmetries in $D^0 \to K^+K^-$ and $D^0 \to \pi^+\pi^-$ represents an estimate of $\beta_{c,eff}$. Due to the small predicted value of $\beta_c$ no experiments can measure this angle, but one can test if the phase difference constrained by experiment is consistent with zero as expected in the standard model. This makes such a measurement appealing when looking for new physics. Furthermore the single decays, and $D^0 \to K^+K^-$ in particular due to larger $BR$, can be used to evaluate the mixing phase $\phi_{MIX}$~\cite{bim}~\cite{inguglia}. It has been estimated that the LHCb Collaboration can constrain $\beta_{c,eff}$ with a statistical precision of $1.9^\circ$ and measure $\phi_{MIX}$ with a statistical precision of $1.8^\circ$ when using a $5~fb^{-1}$ of data, when performing the same measurement with $50~ab^{-1}$ of data collected at Belle II we estimated that a statistical precision of $1.7^\circ$ and $1.8^\circ$ can be achieved on $\beta_{c,eff}$ and $\phi_{MIX}$ respectively~\cite{bim}~\cite{inguglia}.

\section{Comments}
It is clear that searches for $CP$ symmetry breaking in the charm sector have produced many interesting results, however the message here is that since the \textit{charm era} has just began, and one should endeavour to search for $CP$ violation in all possible ways. An advantage of time-dependent measurement is that one measures both the real and the imaginary part of $\lambda_f$ simultaneously, which in turn can be used to simplify understanding the underlying phase structure of the interfering amplitudes being studied. I suggest here that these appealing searches need to proceed and additional complementary methods should be considered from the community. The benefit of such an approach is that we learn about $CP$ violating asymmetries independently of ether requirement to understand strong phase differences manifest in direct $CP$ violation. One can measure the phase of mixing using this methodology, and can constrain $\beta_{c,eff}$ by performing a null test of the phase difference between neutral $D$ decays to $KK$ and $\pi\pi$ final states.

\Acknowledgements
Participation to this workshop has been possible thanks to partial expenses covering offered by the organizing committee and thanks to Queen Mary, University of London.


\begin{thebibliography}{99}


\bibitem{belle}
M. Staric et al.(Belle Collaboration), Phys.Lett.B \boldmath{670} 670:190-195, 2008, arXiv: 0807.0148.
\bibitem{cdf}
CDF note 10784, 2012.
\bibitem{belle1}
B. R. Ko (Belle Collaboration), arXiv: 1212.1975v1.
\bibitem{babar}
B. Aubert et al. (\babar Collaboration), Phys.Rev.Lett. \boldmath{100} (2008) 061803, arXiv: 0709.2715.
\bibitem{lhcbold}
LHCb Collaboration, 	Phys.Rev.Lett. \boldmath{108} (2012) 111602, arXiv: 1112.0938. 
\bibitem{lhcbmuon}
LHCb Collaboration,	Phys. Lett. B \boldmath{723} (2013) 33-43, arXiv: 1303.2614.
\bibitem{lhcbnew}
LHCb Collaboration, LHC\boldmath{b}-CONF-2013-003 (2013).
\bibitem{hfag}
HFAG, (\textit{heavy flavour averaging group web pages, 2012-2013}).
\bibitem{lhcbpic}
LHCb public wep: http://lhcb-public.web.cern.ch (2013). 
\bibitem{bim1}
A. Bevan, G. Inguglia, B. Meadows (2013), \textit{to appear soon}.
\bibitem{bim}
A. Bevan, G. Inguglia, B. Meadows (2011), Phys. Rev. D {\bf 84}, 114009, arXiv:1106.5075v2.
\bibitem{inguglia}
G. Inguglia, Il Nuovo Cimento C, DOI: 10.1393/ncc/i2012-11374-6, pp. 389-398 (2012), arXiv:1204.2303.



\end{thebibliography}
\end{document}